\documentclass[12pt]{article} 
\usepackage{cite}
\usepackage{wrapfig}
\usepackage{amssymb}
\usepackage{amsfonts}
\usepackage{authblk}
\usepackage{amsmath}
\usepackage{lineno}
\usepackage{color}
\usepackage{xcolor}
\usepackage{multirow}
\usepackage{longtable}
\usepackage{lscape}
\usepackage[T2A]{fontenc}
\usepackage[utf8x]{inputenc}
\usepackage{hyperref}
\usepackage{bigstrut}

\usepackage{ifpdf}
\ifpdf
  \usepackage[pdftex]{graphicx}
  \usepackage{epstopdf}
\else
  \usepackage{graphicx}
\fi
\DeclareGraphicsExtensions{.png,.pdf,.jpg,.mps,.eps}
\graphicspath{{./figs/}}

\begin{document}

\title{The second-class current decays $\tau \to \pi \eta(\eta') \nu_{\tau}$ in the NJL model 
including the interaction of mesons in the final state}

\author[1]{M. K. Volkov \footnote{volkov@theor.jinr.ru}}
\author[1,2,3]{K. Nurlan \footnote{nurlan@theor.jinr.ru}}
\author[1]{A.A. Pivovarov \footnote{tex$\_$k@mail.ru}}

\affil[1]{\small Bogoliubov Laboratory of Theoretical Physics, JINR, 
                 141980 Dubna, Moscow region, Russia}
\affil[2]{\small The Institute of Nuclear Physics, Ministry of Energy of 
                  the Republic of Kazakhstan, Almaty, 050032, Kazakhstan}
\affil[3]{\small Al-Farabi Kazakh National University, 
                  Almaty, 050040 Kazakhstan}
                  
\date{}

\maketitle

\abstract{
The effect of the interaction of mesons in the final state is additionally considered 
within the description of $\tau \to \pi \eta(\eta') \nu_{\tau}$ decays. This interaction 
is taken into account at the level of production of intermediate pions. One of them, 
in turn, might be transited into $\eta$ or $\eta'$ mesons. Our results do not exceed 
the experimentally established branching fractions, and they are in agreement 
with the results of other theoretical studies.
}

\section{Introduction}
The hadronic $\tau$ lepton decays play an important role in the study of the intrinsic 
properties of the fundamental QCD theory. The decays $\tau \to \pi \eta(\eta') \nu_{\tau}$ 
are interesting examples of the second class current decays. For the first time, these 
types of currents were considered by Weinberg \cite{Weinberg:1958ut}. These decays 
are suppressed by G-parity violation and can occur only due to the mass difference 
between light $u$ and $d$ quarks. This difference also ensures the transition of pions 
into $\eta$ and $\eta'$ mesons. 

Early experiments to study the $\tau \to \pi \eta \nu_{\tau}$ decay were carried out with 
the CLEO \cite{Artuso:1992qu, Bartelt:1996iv} and ALEPH \cite{Buskulic:1996qs} detectors. 
However, stricter limits for branching fractions with a large number of events are set in 
the experiments of the BaBar \cite{Aubert:2008nj, delAmoSanchez:2010pc, Lees:2012ks} 
$Br(\tau \to \pi \eta \nu_{\tau}) <9.9 \times 10^{-5} $, $Br(\tau \to \pi \eta' \nu_{\tau})<4.0 \times 10^{-6}$ 
and the Belle \cite{Inami:2008ar, Hayasaka:2009zz} $Br(\tau \to \pi \eta \nu_{\tau}) <7.3 \times 10^{-5}$ 
collaborations at $e^+e^-$ colliders. Also, in the recent work \cite{Ogawa:2020iwi}, on the 
base of analysing Belle data the authors presented the branching fractions 
$Br(\tau \to \pi \eta \nu_{\tau})= 4.4 \times 10^{-5} $. From a theoretical point of view, these 
processes were investigated within the framework of various phenomenological models 
\cite{Bramon:1987zb, Neufeld:1994eg, Nussinov:2008gx, Nussinov:2009sn, Paver:2010mz, Paver:2011md, Descotes-Genon:2014tla, Escribano:2016ntp, Hernandez-Tome:2017pdc, Garces:2017jpz}. 

At the same time, in the paper \cite{Volkov:2012be}, the processes $\tau \to \pi \eta(\eta') \nu_{\tau}$ 
are described in the framework of the standard and extended Nambu -- Jona-Lasinio (NJL) model. 
The standard NJL model describes 4 meson nonets in the ground state and their interactions 
within the $U(3)\times U(3)$ chiral symmetry 
\cite{Volkov:1986zb, Ebert:1985kz, Vogl:1991qt, Klevansky:1992qe, Volkov:1993jw}. The extended 
model includes intermediate mesons in the both ground and first radially excited states without 
breaking of the $U(3) \times U(3)$ chiral symmetry 
\cite{Volkov:1996br, Volkov:1996fk, Volkov:2005kw, Volkov:2017arr}. 

In recent years, papers has been published where the effect of interaction of mesons in 
the final state (FSI) considered by taking into account the meson loops describing the 
exchange of a vector meson between outgoing pseudoscalar mesons. These are, for 
example, such decays as $\tau \to [\pi\pi, K \pi, K \eta] \nu_{\tau}$ \cite{Volkov:2020dvz, Kpi, Volkov:2021sma}. 
After that, it turned out to be correct implementation of the generated mesons interaction 
in the previously considered processes $\tau \to \pi \eta(\eta') \nu_{\tau}$ \cite{Volkov:2012be}. 
For this, we can use the recently obtained results in the description of the $\tau \to \pi \pi \nu_{\tau}$ 
\cite{Volkov:2020dvz} decay with considering FSI. Herewith, one should take into 
account the transitions $\pi^0 - \eta (\eta')$ in the final state. Thus, FSI will be taken 
into account here at the level of intermediate pions. Our results are in satisfactory 
agreement with the experimental limitations, as well as in qualitative agreement with 
the estimates obtained by previous authors in other theoretical models. 

It should be noted that when describing FSI using meson loops, we had to go beyond 
the limitations in which the NJL model was formulated. Namely, we consider a higher 
order in $1/N_C$ in perturbation theory, where $N_C$ is the number of colors in QCD.

\section{The decay $\tau \to \pi \eta \nu_{\tau}$ including FSI}

A description of the $\tau \to \pi \pi \nu_{\tau}$ decay with consider of the FSI effect was 
obtained in the paper \cite{Volkov:2020dvz} with satisfactory agreement with the current 
experimental data. For the description of the $\tau \to \pi \eta \nu_{\tau}$ decay, we use 
the amplitude from \cite{Volkov:2020dvz} with an additional $\pi^{0} - \eta$ transition in 
the final state. The corresponding Feynman diagrams are presented in Figures 1-2. As 
a result, for the amplitude of the vector channel of the $\tau \to \pi \eta \nu_{\tau}$ decay 
we obtain 

\begin{eqnarray}
\mathcal{M}(\tau \to \pi \eta \nu_{\tau}) = G_F V_{ud} T_{\pi \eta} l_{\mu} M^2_{\rho} 
\left(1-\frac{i\sqrt{s}\Gamma_{\rho}}{M^2_{\rho}} \right) BW_{\rho} 
\left[ (p_{\eta}-p_{\pi})^{\mu}+ g^2_{\rho} \left( a(s) p^{\mu}_{\eta} -b(s) p^{\mu}_{\pi} \right)\right].
\end{eqnarray} 
Here $G_{F}$ is the Fermi constant, $V_{ud}$ is the element of the 
Cabibbo-Kobayashi-Maskawa matrix; $l_{\mu}$ is the lepton current; $g_{\rho}$ is the 
$\rho \to \pi^+ \pi^-$ decay constant \cite{Volkov:1986zb, Volkov:2020dvz}; $s = (p_{\eta} + p_{\pi})^2$; 
$M_{\rho}= 775.26 \pm 0.23$ MeV, $\Gamma_{\rho}=149.1 \pm 0.8$ MeV are $\rho$ meson 
mass and width \cite{ParticleDataGroup:2020ssz}. The intermediate $\rho$ meson is 
described by the Breit-Wigner propagator \cite{Volkov:2020dvz}. The multipliers $a(s)$ 
and $b(s)$ in the momenta of outgoing particles appear as the functions of $s$ due to FSI. 
These functions read
\begin{eqnarray}
\label{afunc}
a(s) = \frac{I_{\rho}}{M^2_{\rho}} + I_{\rho\pi}+I_{\rho 2\pi}\frac{M^2_{\pi}(M^2_{\eta} - M^2_{\pi})}{M^2_{\rho}} 
- I_{\rho3\pi}\frac{M^4_{\pi}( -M^2_{\eta}+7M^2_{\pi}+6M^2_{\rho}+s)}{6M^2_{\rho}} \\ \nonumber
- I_{\rho4\pi}\frac{M^6_{\pi}( 23M^2_{\eta}+M^2_{\pi}-5s)}{6M^2_{\rho}} 
+ 4 I_{\rho5\pi}\frac{M^8_{\pi}( 4M^2_{\eta}+2M^2_{\pi}-s)}{6M^2_{\rho}},
\end{eqnarray} 
\begin{eqnarray}
\label{bfunc}
b(s) = \frac{I_{\rho}}{M^2_{\rho}} + I_{\rho\pi}\frac{M^2_{\eta}-M^2_{\pi}+M^2_{\rho}}{M^2_{\rho}}
- I_{\rho3\pi}\frac{M^4_{\pi}(13M^2_{\eta}-7M^2_{\pi}+6M^2_{\rho}+s)}{6M^2_{\rho}} \\ \nonumber
- I_{\rho4\pi}\frac{M^6_{\pi}(M^2_{\eta}+23 M^2_{\pi}-5s)}{6M^2_{\rho}} 
+ 4 I_{\rho5\pi}\frac{M^8_{\pi}( 2M^2_{\eta}+4M^2_{\pi}-s)}{6M^2_{\rho}},
\end{eqnarray} 
where $M_{\pi}=139.57$ MeV, $M_{\eta}=547.86 \pm 0.01$ MeV are the meson masses 
\cite{ParticleDataGroup:2020ssz}. If we make the substitution $M^2_{\eta} \to M^2_{\pi}$, 
these expressions will correspond to the similar expressions obtained for the decay 
$\tau \to \pi\pi \nu_{\tau}$ \cite{Volkov:2020dvz}. Integrals over meson loops have the form 
\begin{eqnarray}
\label{integral}
&& I_{\rho n \pi} = \frac{-i}{(2\pi)^{4}}\int\frac{\Theta(\Lambda_{M}^{2} + k^2)}
{(M^2_{\rho} - k^2)({M^2_{\pi}} - k^2)^n}\mathrm{d}^{4}k,
\end{eqnarray}
the value of the cutoff parameter is $\Lambda_{M}=760$ MeV \cite{Volkov:2020dvz}.
Nondiagonal $\pi^0 - \eta(\eta')$ transitions can occur only due to the difference between 
the masses of light $u$ and $d$ quarks. The amplitudes of these transitions in the NJL 
quark model take the form \cite{Volkov:2012be}
\begin{eqnarray}
T_{\pi\eta(\eta')} = 2g^2_{\pi} \left[ \left(2I^{m_d}_1 + M^2_{\eta(\eta')}I^{m_d}_2 \right) - \left( 2I^{m_u}_1 + M^2_{\eta(\eta')}I^{m_u}_2 \right) \right] \frac{\sin(\bar{\theta})(\cos(\bar{\theta})) }{M^2_{\pi}-M^2_{\eta(\eta')}}, 
\end{eqnarray} 
where $m_u = 280$ MeV and $m_d = 283.7$ MeV are the constituent quark masses \cite{Volkov:1986zb}; 
$\theta= -54^{\circ}$ is the mixing angle of $\eta$ and $\eta'$ mesons \cite{Volkov:1998ax}; 
$g_{\pi}={m_u}/F_{\pi}$ is the $\pi \to \mu \nu_{\mu}$ decay constant \cite{Volkov:1986zb}. The 
comparison of the amplitude $T_{\pi\eta(\eta')}$ and coupling constant  $g_{\rho\pi\eta(\eta')}$ is 
given in the Table 1.  

\begin{table}[h!]
\label{Tabddd}
\begin{center}
\begin{tabular}{cccc}
\hline
   & NJL model & \cite{Paver:2010mz, Paver:2011md} & \cite{Escribano:2016ntp} \\
\hline
$T_{\pi\eta}$ & $1.55 \times 10^{-2}$ & $1.34 \times 10^{-2}$ & $(0.98 \pm 0.03) \times10^{-2}$  \\
$T_{\pi\eta'}$ & $6.80 \times 10^{-3}$ & $(3 \pm 1) \times10^{-3}$ & $(0.25 \pm 0.14) \times 10^{-3}$ \\
\hline
   & NJL model & \cite{Paver:2010mz, Paver:2011md} & \cite{Nussinov:2008gx, Nussinov:2009sn} \\
\hline
$g_{\rho\pi\eta}$ & $9.51 \times 10^{-2}$ & $8.04 \times 10^{-2}$ & $8.50 \times10^{-2}$  \\
$g_{\rho\pi\eta'}$ & $4.17 \times 10^{-2}$ & $(1.8 \pm 0.6) \times10^{-2}$ & $ <2.5 \times 10^{-2}$ \\
\hline
\end{tabular}
\end{center}
\caption{The comparison of $T_{\pi\eta(\eta')}$ and coupling constant $g_{\rho\pi\eta(\eta')}$.}
\end{table}

\begin{figure}[h]
\label{feyn1}
\center{\includegraphics[scale = 0.5]{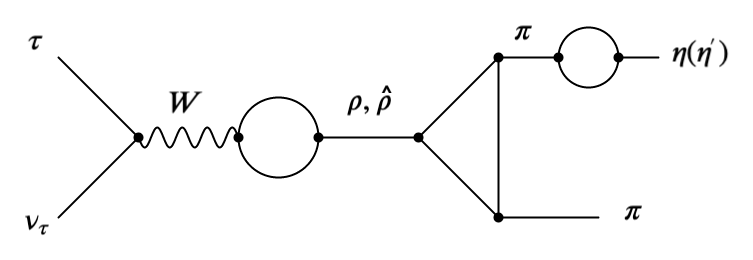}}
\caption{The quark diagram of the decays $\tau \to \pi \eta(\eta') \nu_{\tau}$ in the NJL model}
\end{figure}
\begin{figure}[h]
\label{feyn2}
\center{\includegraphics[scale = 0.4]{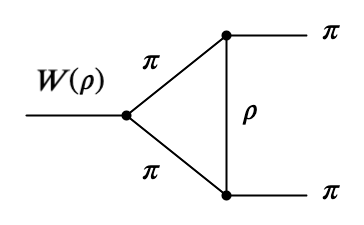}}
\caption{The meson loop diagram describing FSI at the intermediate pion level}
\end{figure}

We can calculate the decay width of the studied decay $\tau \to \pi \eta \nu_{\tau}$ by the formula
\begin{eqnarray}
\Gamma (\tau \to \pi \eta \nu_{\tau}) = \frac{1}{2} \cdot \frac{1}{256 {\pi}^3 M^3_{\tau}} \int\limits_{s_{-}}^{s_{+}}ds \int\limits_{t_{-}(s)}^{t_{+}(s)} dt \: {|\mathcal{M}|}^2,
\end{eqnarray}
where the variables are defined as  $s = (p_{\tau}-p_{\nu})^2=(p_{\eta} + p_{\pi})^2$, 
$t=(p_{\tau} - p_{\eta})^2=(p_{\pi} + p_{\nu})^2$. The limits of integration have the form 
\begin{eqnarray}
s_{+} = M^2_{\tau}, \quad s_{-} = (M_{\eta}+M_{\pi})^2,
\end{eqnarray}
 \begin{eqnarray}
&& t_{\pm}(s) = \frac{1}{2} \biggl[ M^2_{\tau} + M^2_{\eta} + M^2_{\pi} - s + 
\frac{M^2_{\tau}}{s}(M^2_{\eta}-M^2_{\pi}) \pm \sqrt{D(s)} \biggr],
\end{eqnarray}
where
 \begin{eqnarray}
 D(s) =  s^{-2} \left(s-M^2_{\tau}\right)\left(s-(M_{\eta}+M_{\pi})^2\right) 
\left(s^2-s\left(M^2_{\tau}+(M_{\pi}-M_{\eta})^2 \right)+M^2_{\tau}(M_{\pi}-M_{\eta})^2 \right). 
\end{eqnarray} 

The obtained results and the experimentally measured values for the branching fractions 
of the $\tau \to \pi \eta \nu_{\tau}$ process are given in Table 2. In the table comparisons 
with the results of other theoretical studies are also given. In our calculations, we did not 
take into account the FSI effect in the scalar channel since contribution is an order of 
magnitude less than in the vector channel. Moreover, the Lagrangian of the interaction 
of the $a_0$ meson with the pion and the $\eta$ meson in the minimum order does not 
contain derivatives \cite{Volkov:2020dvz}. 

\section{The decay $\tau \to \pi \eta' \nu_{\tau}$ in the extended NJL model including FSI} 

To describe the $\tau \to \pi \eta' \nu_{\tau}$ decay due to a higher threshold for the final 
meson products, it is necessary to take into account the first radially excited states as 
intermediate ones. Therefore, we will use the extended NIL model  
~\cite{Volkov:1996br, Volkov:1996fk, Volkov:2005kw, Volkov:2017arr}. A fragment of the 
chiral quark-meson Lagrangian of the extended NJL model containing the mesons 
involved in the process under consideration has the form
\begin{eqnarray}
\label{lagrangian}
\Delta L_{int} = \bar{q} \left[ \frac{1}{2} \gamma^{\mu} \sum_{j = \pm}\lambda_{j}^{\rho} 
\left(A_{\rho}\rho^{j}_{\mu} + B_{\rho}\hat{\rho}^{j}_{\mu}\right) 
+ i \gamma^{5} \sum_{j = \pm,0} \lambda_{j}^{\pi} \left(A_{\pi}\pi^{j} + 
B_{\pi}\hat{\pi}^{j}\right) \right. \nonumber\\  \left. 
+ i \gamma^{5} \sum_{j = u,s} \lambda_{j}^{\eta} A_{\eta^{j}}\eta'\right]q,
\end{eqnarray}
where $q$ and $\bar{q}$ are u, d and s quark fields with constituent quark masses 
$m_{u} \approx m_{d} = 280$~MeV, $m_{s} = 420$~MeV, excited mesonic states are 
marked with a hat, $\lambda$ are linear combinations of the Gell-Mann matrices 
\cite{Volkov:2017arr},
\begin{eqnarray}
\label{verteces1}
	A_{M} = \frac{1}{\sin(2\theta_{M}^{0})}\left[g_{M}\sin(\theta_{M} + \theta_{M}^{0}) +
	g'_{M}f_{M}(k_{\perp}^{2})\sin(\theta_{M} - \theta_{M}^{0})\right], \nonumber\\
	B_{M} = \frac{-1}{\sin(2\theta_{M}^{0})}\left[g_{M}\cos(\theta_{M} + \theta_{M}^{0}) +
	g'_{M}f_{M}(k_{\perp}^{2})\cos(\theta_{M} - \theta_{M}^{0})\right].
\end{eqnarray}
The subscript M indicates the corresponding meson; $\theta_{\pi} = 59.48^{\circ}, 
\theta_{\rho} = 81.8^{\circ}, \theta_{\pi}^{0} = 59.12^{\circ}, \theta_{\rho}^{0} = 61.5^{\circ}$ are 
the mixing angles \cite{Volkov:2017arr}.

For the $\eta'$ meson, the factor $A$ takes a slightly different form. This is due to the fact that 
in the case of the $\eta'$ meson four states are mixed \cite{Volkov:2017arr}:
\begin{eqnarray}
\label{verteces2}
	A_{\eta^{u}} = -0.32 g_{\eta^{u}} - 0.48 g'_{\eta^{u}} f_{uu}(k_{\perp}^{2}), \nonumber\\
	A_{\eta^{s}} = 0.56 g_{\eta^{s}} + 0.3 g'_{\eta^{s}} f_{ss}(k_{\perp}^{2}).
\end{eqnarray}
Here $f\left(k_{\perp}^{2}\right) = \left(1 + d k_{\perp}^{2}\right)\Theta(\Lambda^{2} - k_{\perp}^2)$ is 
the form-factor describing the first radially excited meson states. The slope parameters, 
$d_{uu} = -1.784 \times 10^{-6} \textrm{MeV}^{-2}$ and $d_{ss} = -1.737 \times 10^{-6} \textrm{MeV}^{-2}$, 
are unambiguously fixed from the condition of constancy of the quark condensate after the inclusion 
of radially excited states and depends only on the quark composition of the corresponding meson.

The quark-meson coupling constants have the form
\begin{eqnarray}
	\label{Couplings}
g_{\rho} = \left(\frac{2}{3}I_{20}\right)^{-1/2}, \, g'_{\rho} = \left(\frac{2}{3}I_{20}^{f^{2}}\right)^{-1/2}, 
\, g_{\pi} =  \left(\frac{4}{Z_{\pi}}I_{20}\right)^{-1/2}, \, g'_{\pi} =  \left(4I_{20}^{f^{2}}\right)^{-1/2}, \nonumber\\
\quad g_{\eta^{u}} = \left(\frac{4}{Z_{\eta^{u}}}I_{20}\right)^{-1/2}, \, g'_{\eta^{u}} =  \left(4 I_{20}^{f^{2}}\right)^{-1/2}, 
\, g_{\eta^{s}} =  \left(\frac{4}{Z_{\eta^{s}}}I_{02}\right)^{-1/2}, \, g'_{\eta^{s}} =  \left(4 I_{02}^{f^{2}}\right)^{-1/2}, 
\end{eqnarray}
here $Z_{\pi} \approx Z_{\eta^{u}}$ and $Z_{\eta^{s}}$ are additional renormalization constants appearing 
in the $\pi-a_{1}$ and $\eta-f_{1}(1420)$ transitions \cite{Volkov:2017arr}.

Integrals appearing in the quark loops are
\begin{eqnarray}
	I_{n_{1}n_{2}}^{f^{m}} =
	-i\frac{N_{c}}{(2\pi)^{4}}\int\frac{f^{m}(k^2_{\perp})}{(m_{u}^{2} - k^2)^{n_{1}}(m_{s}^{2} - k^2)^{n_{2}}}\Theta(\Lambda_{3}^{2} - k^2_{\perp})
	\mathrm{d}^{4}k,
\end{eqnarray}
where $\Lambda_3=1030$ MeV is the cutoff parameter \cite{Volkov:2017arr}.

Using the extended NJL model for the $\tau \to \pi \eta' \nu_{\tau}$ process, after taking into account FSI, 
we obtain the following amplitude: 
\begin{eqnarray}
\label{amplitude1}
\mathcal{M}(\tau \to \pi \eta' \nu_{\tau}) = G_{F} V_{ud} Z_{\pi} T_{\pi\eta'} l_{\mu} \left\{ \left[\mathcal{M}_{c} + \mathcal{M}_{\rho} + \mathcal{M}_{\hat{\rho}}\right]^{\mu\nu} \left(p_{\eta} - p_{\pi}\right)_{\nu} \right. \nonumber\\ 
 \left. + \left[\mathcal{M}_{c(loop)} + \mathcal{M}_{\rho(loop)} + \mathcal{M}_{\hat{\rho}(loop)}\right]^{\mu\nu}\left(a(s) p_{\eta} - b(s) p_{\pi}\right)_{\nu} \right\},
\end{eqnarray}
here the functions $a(s)$ and $b(s)$ are obtained by replacing $M^2_{\eta} \to M^2_{\eta'}$ in accordance 
with the definitions (\ref{afunc}) and (\ref{bfunc}). The terms in the square brackets (\ref{amplitude1}) 
describe the contributions from the contact diagram and diagrams with intermediate $\rho$, $\hat{\rho}$ 
mesons in the ground and first radially excited states: 
\begin{eqnarray}
    \mathcal{M}_{c}^{\mu\nu} & = & \left[1 - C^2_{\rho} \frac{6 m_{u}^{2}}{M_{a_{1}}^{2}}\right] g^{\mu\nu}, \nonumber
\end{eqnarray}
\begin{eqnarray}
    \mathcal{M}_{\rho}^{\mu\nu} & = & C_{\rho}^{2} \left[1 - 4 I_{20}^{\rho a_{1}} \frac{m_{u}^{2}}{M_{a_{1}}^{2}}\right] \frac{g^{\mu\nu}p^{2} - p^{\mu}p^{\nu}}{M_{\rho}^{2} - p^{2} - i \sqrt{p^{2}}\Gamma_{\rho}}, \nonumber
\end{eqnarray}  
\begin{eqnarray}
    \mathcal{M}_{\hat{\rho}}^{\mu\nu} & = & e^{i \pi} C_{\hat{\rho}}^{2} \left[1 - 4 I_{20}^{\hat{\rho} a_{1}} \frac{C_{\rho}}{C_{\hat{\rho}}} \frac{m_{u}^{2}}{M_{a_{1}}^{2}}\right] \frac{g^{\mu\nu}p^{2} - p^{\mu}p^{\nu}}{M_{\hat{\rho}}^{2} - p^{2} - i \sqrt{p^{2}}\Gamma_{\hat{\rho}}}, \nonumber
\end{eqnarray}  
\begin{eqnarray}    
    \mathcal{M}_{c(loop)}^{\mu\nu} & = & g_{\rho}^{2} Z_{\pi}^{2} C_{\rho}^{2} \left[1 - C_{\rho}^{2} \frac{6 m_{u}^{2}}{M_{a_{1}}^{2}}\right]\left[1 - 4 I_{20}^{\rho a_{1}} \frac{m_{u}^{2}}{M_{a_{1}}^{2}}\right]^{2} g^{\mu\nu}, \nonumber
\end{eqnarray}    
\begin{eqnarray}     
    \mathcal{M}_{\rho(loop)}^{\mu\nu} & = & g_{\rho}^{2} Z_{\pi}^{2} C_{\rho}^{4} \left[1 - 4 I_{20}^{\rho a_{1}} \frac{m_{u}^{2}}{M_{a_{1}}^{2}}\right]^{3} \frac{g^{\mu\nu}p^{2} - q^{\mu}q^{\nu}}{M_{\rho}^{2} - p^{2} - i \sqrt{p^{2}}\Gamma_{\rho}}, \nonumber
\end{eqnarray} 
\begin{eqnarray}      
    \mathcal{M}_{\hat{\rho}(loop)}^{\mu\nu} & = & e^{i \pi} g_{\rho}^{2} Z_{\pi}^{2} C_{\hat{\rho}}^{2} C_{\rho}^{2} \left[1 - 4 I_{20}^{\hat{\rho} a_{1}} \frac{C_{\rho}}{C_{\hat{\rho}}} \frac{m_{u}^{2}}{M_{a_{1}}^{2}}\right]\left[1 - 4 I_{20}^{\rho a_{1}} \frac{m_{u}^{2}}{M_{a_{1}}^{2}}\right]^{2} \frac{g^{\mu\nu}p^{2} - p^{\mu}p^{\nu}}{M_{\hat{\rho}}^{2} - p^{2} - i \sqrt{p^{2}}\Gamma_{\hat{\rho}}}, \nonumber
\end{eqnarray}
where $M_{a_1}=1230 \pm 40$ MeV, $M_{\rho'}=1465 \pm 25$ MeV, $\Gamma_{\rho'}=400 \pm 60$ MeV 
are masses and width of $a_1$ and $\rho'$ mesons, respectively, given in PDG \cite{ParticleDataGroup:2020ssz}.
Following the paper \cite{Volkov:2012be}, for the excited states we use the phase factor~$e^{i \pi}$. The 
constants $C_{\rho} \approx 0.95$ and $C_{\hat{\rho}} \approx 0.31$ are taken from \cite{Volkov:2017arr}. 

Integrals with the vertices from the Lagrangian (\ref{lagrangian}) in the numerator, which were also 
used in the amplitude, take the form: 
	\begin{eqnarray}
	\label{DiffIntegral}
		I_{n_{1} n_{2}}^{M, \dots,\hat{M}, \dots} & = &
		-i\frac{N_{c}}{(2\pi)^{4}}\int\frac{A_{M} \dots B_{M} \dots}{(k^2 - m_{u}^{2})^{n_{1}}(k^2 - m_{s}^{2})^{n_{2}}} \Theta(\Lambda_{3}^{2} - k^2_{\perp})	\mathrm{d}^{4}k,
	\end{eqnarray}
	where $A_{M}, B_{M}$ are defined in (\ref{verteces1}).

The results obtained using the amplitude (\ref{amplitude1}) are presented in Table 2.

\begin{table}[h!]
\label{Tabddd}
\begin{center}
\begin{tabular}{ccc}
\hline
   & Br$(\tau \to \pi \eta \nu_{\tau}) \times 10^5$ & Br$(\tau \to \pi \eta' \nu_{\tau}) \times 10^7$  \\
\hline
NJL model & $1.69$ & $1.17$ \\
\cite{Neufeld:1994eg} & 1.21 & ---  \\
\cite{Nussinov:2008gx, Nussinov:2009sn} & 1.36 & 2 -- 14 \\
\cite{Paver:2010mz, Paver:2011md} & $0.4-2.9$ & $0.6-2.1$ \\
\cite{Descotes-Genon:2014tla} & $0.33$ & --- \\
\hline
BaBar \cite{delAmoSanchez:2010pc, Lees:2012ks} & $<9.9$ & $<40$ \\
Belle \cite{Hayasaka:2009zz} & $<7.3$ & --- \\
\hline
\end{tabular}
\end{center}
\caption{The comparison of the branching fractions for the decays $\tau \to \pi\eta(\eta')\nu_{\tau}$.}
\end{table}

It worth to note, that the similar results for the branching fraction were obtained in the work 
\cite{Neufeld:1994eg} using the chiral perturbation theory (ChPT) with resonances, and in 
\cite{Nussinov:2008gx, Nussinov:2009sn} using the non-standard $V-A$ scalar weak interaction. 
The ChPT model is close to our model since it is also based on chiral symmetries of strong 
interactions. Also, the results obtained in the papers \cite{Paver:2010mz, Paver:2011md} using 
the vector dominance do not contradict our results. The decay widths calculated in this work 
overestimates the calculated results in the paper \cite{Descotes-Genon:2014tla}. The reason 
might be in quite different parameterization of the form factors. 

The theoretical uncertainty of the NJL model can be estimated at $ \sim 5 \%$. The source of 
this uncertainty is the effects of chiral symmetry breaking \cite{Volkov:2017arr}. 

\section{Conclusions}

In recent works \cite{Volkov:2020dvz, Kpi}, it was shown that taking FSI into account plays an 
important role in describing the $\tau$ decays with pseudoscalar mesons in the final state. In 
the present paper, we have confirmed this by taking into account the corresponding corrections 
to the previously described $\tau \to \pi \eta(\eta')\nu_\tau$ decays. Furthermore, our results 
do not contradict the experimental data, and are in agreement with the number of authors 
\cite{Neufeld:1994eg, Nussinov:2008gx, Nussinov:2009sn}. In \cite{Escribano:2016ntp}, these 
processes are also studied within the framework of the chiral perturbation theory with resonances. 
In this case, the parametrizations of vector and scalar form factors were used. Also, in the 
work \cite{Garces:2017jpz}, possible contributions from New Physics for the studied decays 
were estimated. 

In the above theoretical works, the FSI effect was taken into account by parametrizing the 
corresponding form factors. We took FSI into account due to the exchange of pions by the 
$\rho$ meson in the P wave. Herewith, we had to go beyond the lower order $1/N_{C}$ 
expansion, in which the NJL model was formulated. 

The Belle II experimental collaboration presented the research program in recent works 
\cite{Belle-II:2010dht, Belle-II:2018jsg}. The upcoming experiment will allow to study the 
second class current decays more accurately. We hope that our results will receive 
experimental confirmation. 

\subsection*{Acknowledgments}
We are grateful to A.B. Arbuzov for his interest in our work and important remarks which 
improved the paper; this work is supported by the JINR grant for young scientists and 
specialists No. 21-302-04. The research of K. Nurlan was funded by the Science Committee 
of the Ministry of Education and Science of the Republic of Kazakhstan (Grant No.AP09057862).

\end{document}